\documentclass[a4paper,fleqn,usenatbib]{mnras}

\usepackage{mathtools}  
\usepackage{amssymb}
\usepackage[T1]{fontenc}
\usepackage{ae,aecompl}
\usepackage{adjustbox}

\usepackage{graphicx}	
\usepackage{amsmath}	
\usepackage{amssymb}	
\usepackage{natbib} 

\title[Polarimetry through multi-line observations II]{Chromospheric polarimetry through multi-line observations of the 850~nm spectral region II: A magnetic flux tube scenario}

\author[C. Quintero Noda et al.]{C. Quintero Noda,$^{1}$\thanks{E-mail: carlos@solar.isas.jaxa.jp}
Y. Kato,$^{2,3}$
Y. Katsukawa,$^{4}$
T. Oba,$^{5}$
J. de la Cruz Rodr\'{i}guez,$^{6}$
\newauthor
M. Carlsson,$^{3}$
T. Shimizu,$^{1}$
D. Orozco Su\'arez,$^{7}$
B. Ruiz Cobo,$^{8,9}$
M. Kubo,$^{4}$
\newauthor
T. Anan,$^{10}$
K. Ichimoto,$^{4,10}$
Y. Suematsu$^{4}$ 
\\
$^{1}$Institute of Space and Astronautical Science, Japan Aerospace Exploration Agency, Sagamihara, Kanagawa 252-5210, Japan\\
$^{2}$Department of Physics, Faculty of Science, Chiba University, Inage-ku, Chiba 263-8522, Japan\\
$^{3}$Institute of Theoretical Astrophysics, University of Oslo, P.O. Box 1029 Blindern, N-0315 Oslo, Norway\\
$^{4}$National Astronomical Observatory of Japan, 2-21-1 Osawa, Mitaka, Tokyo 181-8588, Japan\\
$^{5}$SOKENDAI, Shonan Village, Hayama, Kanagawa 240-0193 Japan\\
$^{6}$Institute for Solar Physics, Dept. of Astronomy, Stockholm University, Albanova University Center, SE-10691 Stockholm, Sweden\\
$^{7}$Instituto de Astrof\'isica de Andaluc\'ia (CSIC), Glorieta de la Astronom\'ia, 18008 Granada, Spain\\
$^{8}$Instituto de Astrof\'isica de Canarias, E-38200, La Laguna, Tenerife, Spain.\\
$^{9}$Departamento de Astrof\'isica, Univ. de La Laguna, La Laguna, Tenerife, E-38205, Spain\\
$^{10}$Kwasan and Hida Observatories, Kyoto University, Kurabashira Kamitakara-cho, Takayama-city, 506-1314 Gifu, Japan\\
}

\date{Accepted XXX. Received YYY; in original form ZZZ}

\pubyear{2017}

\begin{document}
\label{firstpage}
\pagerange{\pageref{firstpage}--\pageref{lastpage}}
\maketitle

\begin{abstract}
In this publication we continue the work started in \cite{QuinteroNoda2017} examining this time a numerical simulation of a magnetic flux tube concentration. Our goal is to study if the physical phenomena that take place in it, in particular, the magnetic pumping, leaves a specific imprint on the examined spectral lines. We find that the profiles from the interior of the flux tube are periodically dopplershifted following an oscillation pattern that is also reflected in the amplitude of the circular polarization signals. In addition, we analyse the properties of the Stokes profiles at the edges of the flux tube discovering the presence of linear polarization signals for the Ca~{\sc ii} lines, although they are weak with an amplitude around 0.5\% of the continuum intensity. Finally, we compute the response functions to perturbations in the longitudinal field and we estimate the field strength using the weak field approximation. Our results indicate that the height of formation of the spectral lines changes during the magnetic pumping process which makes the interpretation of the inferred magnetic field strength and its evolution more difficult. These results complement those from previous works demonstrating the capabilities and limitations of the 850~nm spectrum for chromospheric Zeeman polarimetry in a very dynamic and complex atmosphere.
\end{abstract}





\section{Introduction}

The energy that is released into the corona is transported through and modulated by the chromosphere. Magnetic fields greatly influence the structuring and energy balance of this layer and, for this reason, future space missions such as Solar-C \citep{Katsukawa2011,Watanabe2014,Suematsu2016} and ground-based telescopes such as DKIST \citep{Keil2011} or EST \citep{Collados2013} aim to understand the properties of the chromospheric magnetic field through routine polarimetric measurements of spectral lines that form in this layer. 

\begin{figure*}
\begin{center} 
 \includegraphics[trim=36 0 -10 0,width=17.5cm]{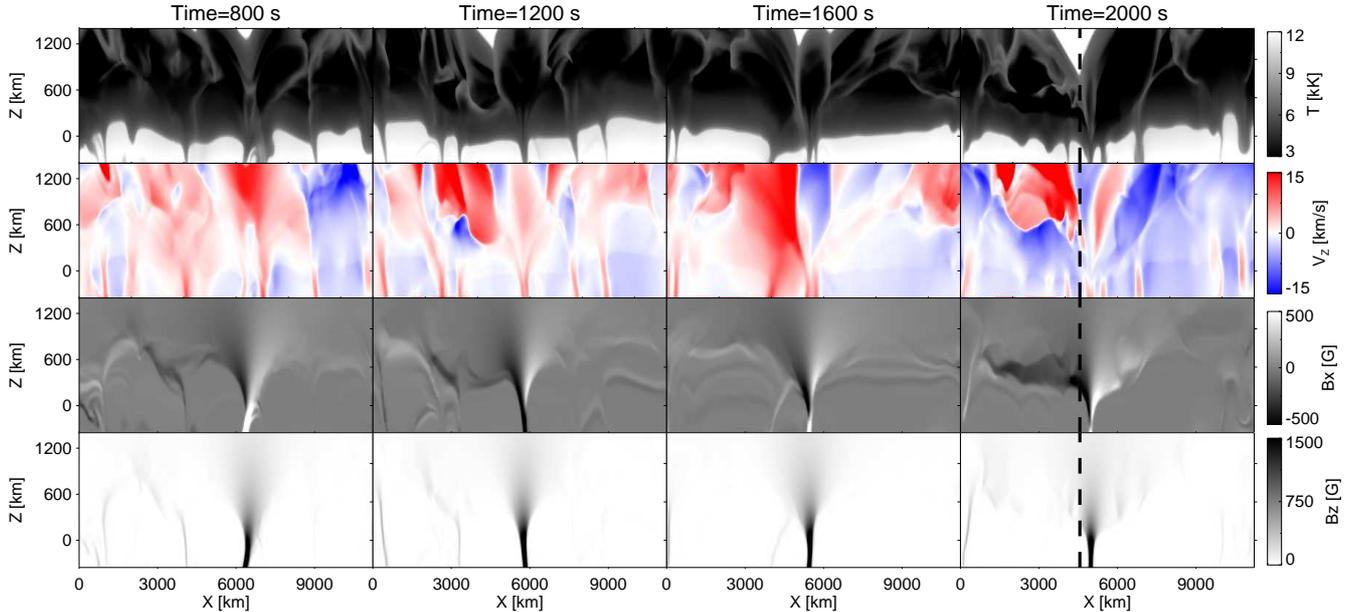}
 \vspace{-1.0cm}
 \caption{From top to bottom, temperature, line-of-sight (LOS) velocity, and the horizontal and vertical components $B_x$ and $B_z$ of the magnetic field, respectively ($B_y=0$). The sign of the LOS velocity follows the traditional spectroscopic convention where negative values designate material moving upwards with  respect to the solar surface. Each column corresponds to a different time. The abscissa depicts the horizontal computational domain while the ordinate displays the geometrical height. The spatial location marked with a dashed line will be examined later.}
 \label{Context}
 \end{center}
\end{figure*}

We studied in \cite{QuinteroNoda2017} the spectral region around 850~nm finding several highly capable lines for photospheric and chromospheric polarimetry, e.g. Fe~{\sc i} 8468~\AA, and Ca~{\sc ii} 8498 and 8542~\AA. We mentioned in that work that those spectral lines can fit on a single camera of a spectro-polarimetric instrument at one time. For instance, on the Sunrise Chromospheric Infrared spectro-Polarimeter (SCIP; Katsukawa et al., in preparation), that aims to cover the mentioned spectrum of around 90~\AA \ in one single channel. This instrument has a spectral resolution of $2\times10^{5}$, a polarization sensitivity of $3\times10^{-4}$ (normalised to the continuum intensity), and a spatial resolution of approximately 0.2 arc sec, corresponding to 150~km on the solar surface. It is currently planned to be on board of the Sunrise balloon-borne telescope \citep{Barthol2011,Berkefeld2011,Gandorfer2011} that, after two successful flights \citep[e.g.,][]{Solanki2010,Solanki2017}, is scheduled for a third one. 

The mentioned spectral window at 850~nm presents itself as one of the most complete spectral regions providing continuous sensitivity to the atmospheric parameters for the continuum optical depth range $\log \tau\sim[0,-5.5]$, which approximately corresponds to the geometrical heights comprehended between $z=[0,1000]$~km above the solar surface. However, we also closed the work presented in \cite{QuinteroNoda2017} explaining that there are several studies that we should perform in order to fully understand the capabilities of the 850~nm spectral region and, most important, its limitations. Among these studies we mentioned the inversion of noisy synthetic profiles from realistic simulations, comparing the advantages of inverting a single chromospheric line, for instance the Ca~{\sc ii} 8542~\AA, versus inverting simultaneously all the spectral lines that fall in the 850~nm window (with different height of formation in the solar atmosphere) or observations with ground-based telescopes of these lines pointing to different magnetic regions. Moreover, we also argued that it could be highly beneficial if we continue examining additional numerical simulations as they bring us the possibility to perform laboratory-like studies where the physical information that produces the synthetic profiles is accurately known in advance.

We focus on the latter case in this work, in particular on the flux sheath simulation presented in \cite{Kato2016}. In this case, we have a 2D simulation comparable to a slit sit-and-stare observation where a highly dynamic flux tube can be found surrounded by a quiet Sun atmosphere. In the following sections we  present a study of how the synthetic profiles respond to the different physical processes that happen in- and outside the flux tube. 

\section{Simulations and methodology}

In this section, we briefly introduce the main physical mechanisms that take place in the simulation described in \cite{Kato2016}. The authors developed that simulation  using the {\sc bifrost} code \citep{Gudiksen2011} that allows extending the vertical domain of the simulation higher in the solar atmosphere up to the lower corona. They started with a snapshot of the atmosphere previously calculated in \cite{Kato2011} containing a single magnetic flux concentration that remains isolated for more than 60 minutes. Although this is a simple configuration, it allows to obtain the basic information related to the thermodynamics inside a flux tube without the complexity that multiple flux concentrations and their interactions bring.

The simulation has a computational domain of $400\times535$ cells that horizontally covers 11.2~Mm while the vertical extent is 12~Mm, being $z=0$ the mean height where the continuum optical depth at 500~nm is unity. The horizontal grid size is constant with a step size of 28~km while the vertical grid size is non-uniform being 13~km at $z=0$~km, and 45~km at the upper boundary in the corona. The time cadence is 2~seconds per output snapshot, and we examine 20~min of the total simulation time. Figure~\ref{Context} displays an example of the vertical stratification of selected physical parameters. We choose four different time snapshots to show the evolution of the magnetic flux concentration. In particular, this structure harbours downflowing material as a consequence of the interaction with the surrounding convective flow and it experiences distortions, as swaying and lateral motions, with time. This interaction excites magneto-acoustic waves within the flux concentration through the so-called magnetic pumping process \citep[see Figure~2 of][]{Kato2016}. 

\subsection{Synthesis of the Stokes profiles}\label{sec22}

Here we summarize the method we use in this work as it is basically the same as in \cite{QuinteroNoda2017}. We synthesise with the {\sc nicole} code \citep{SocasNavarro2000,SocasNavarro2015} the full spectrum shown in Figure~1 of the former publication. We perform column-by-column forward modelling, i.e. each column is treated independently and the non-local thermodynamic equilibrium (non-LTE) atomic populations are solved for assuming a plane-parallel atmosphere. This approximation is valid under LTE conditions and for some strong non-LTE lines, e.g. the Ca~{\sc ii} infrared lines, where the 3D radiation field does not play an important role for the computed population densities  \citep{Leenaarts2009,delaCruzRodriguez2012}.

The synthesised spectral region contains several photospheric lines and two chromospheric lines that belong to the Ca~{\sc ii} infrared triplet. The spectral sampling used is the same as well, i.e. $\Delta\lambda=40$~m\AA.  We perform the synthesis of the Stokes profiles for 20~minutes of simulation, i.e. we synthesise 400 horizontal points at 600 time steps ($400\times600$ pixels). The vertical domain is reduced to $z=[-650,2500]$~km as the spectral lines of interest form completely within this height range. We assume that we are looking at the disc centre, i.e. $\mu=1$ (where $\mu=\cos\theta$, and $\theta$ is the angle of the ray with respect to the normal of the atmosphere).  In addition, no microturbulence is included although, in order to simulate the effect of a general spectral point spread function, we degrade the spectra employing a Gaussian profile with a full width at half maximum of 1.5 km/s, a value similar to that expected for a spectrograph, e.g. Sunrise/SCIP (see the introduction). Finally, we use the original spatial conditions of the simulation, i.e. no spatial degradation.

\subsection{Line core width}

\begin{figure}
\begin{center} 
 \includegraphics[trim=15 0 10 0,width=8.4cm]{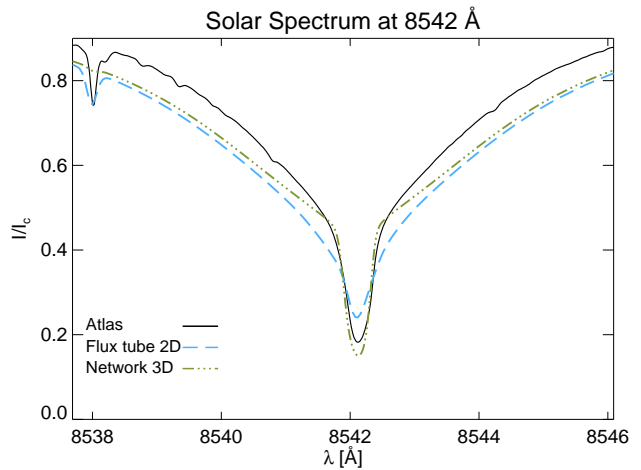}
 \vspace{-0.7cm}
 \caption{Comparison between the solar atlas (solid) and the spatially averaged intensity profile over the entire simulation box (dashed) using a null value for the microturbulence. We add for comparison purposes the spatially averaged intensity profile from the snapshot 385 of the enhanced network simulation (dashed-dotted), computed with a microturbulence value of 3~km/s constant with height.}
 \label{Atlas_compare}
 \end{center}
\end{figure}

Current numerical simulations do not contain sufficient heating and small-scale motions to match the observed intensities and widths of chromospheric lines \citep[for instance,][]{Leenaarts2009}. Moreover, if the line core intensity profile is narrower and deeper than expected, this could induce artificially large polarizations signals \citep{delaCruzRodriguez2012}. Therefore, the first test we perform is computing the spatially averaged intensity profile for the whole simulation box presented in Figure~\ref{Context}. The results using a null microturbulence value are depicted by the dashed line in Figure~\ref{Atlas_compare}, while the solid profile corresponds to the solar atlas \citep{Delbouille1973}, and the dashed-dotted line displays the results from the enhanced network simulation \citep{Carlsson2016}, used in the first paper of these series. The intensity profile produced by the 2D simulation (dashed) is wide, showing a line core width at full width half maximum of 535~m\AA, very close to the value displayed by the solar atlas, i.e. 574 m\AA, and the results of additional studies \citep[for instance, ][ obtained a line core width ranging between 450-550~m\AA]{Cauzzi2009}. Therefore, in this case, there is no need of introducing an additional microturbulence contribution. We believe that there are several reasons for this, as the higher spatial resolution of the simulation, i.e. 28~km, but also that the flux sheath (harbouring a highly dynamic plasma) occupies a considerable part of the simulation domain. The latter could also explain the asymmetry of the intensity profile towards the red \citep[we do not include in this work the effect of the isotopic splitting,] []{Leenaarts2014}. An additional factor could be the fact that the mean effective temperature of this simulation is higher than that of \cite{Carlsson2016} and displays larger fluctuations with time.

\subsection{Flux tube}

We focus most of this study on the phenomena that takes place inside the flux tube, although we also examine outer locations later. Therefore, we need to establish a criterion that defines what pixels we consider as pertaining to the magnetic concentration. In this regard, we choose the pixels that fulfil $B_z>1500$~G at the geometrical height of $z=0$~km. The results are presented in Figure~\ref{Pixels}. We can see that highlighted pixels perfectly match at any time the strong concentration of magnetic field at $z=0$~km (top row, rightmost panel) and they are also located in a cool region that harbours strong down-flowing material (left and middle panels). However, as the flux tube oscillates and displays swaying motions with time, we also verify that choosing those pixels, we are still inside the flux tube at higher atmospheric layers. In this regard, we check the geometrical height $z=1000$~km as it is closer to the expected formation height of the Ca~{\sc ii} lines \citep[e.g., see Figure~5 of][]{Cauzzi2008}. We plot in the bottom row of Figure \ref{Pixels} the results and we can see that, again, highlighted pixels are on top of the magnetic flux concentration (rightmost panel) although, at this height, the flux tube is wider. Moreover, we can distinguish the imprint of plasma motions as an oscillatory pattern (the vertical axis represents the simulation time) in the temperature and the line-of-sight (LOS) velocity. For the latter parameter, down-flows and up-flows take place periodically while, in the former, hot patches appear when a change in the LOS velocity takes place, i.e. the propagation of a shock wave preceded by a rarefaction wave of down-flowing material \citep{Kato2016}. Our aim is to study in detail the pixels highlighted with orange/black in Figure~\ref{Pixels} in the following sections.

\section{Results}

\subsection{Stokes profiles inside the flux tube}\label{secshift}

\begin{figure}
\begin{center} 
 \includegraphics[trim=-12 0 15 0,width=8.0cm]{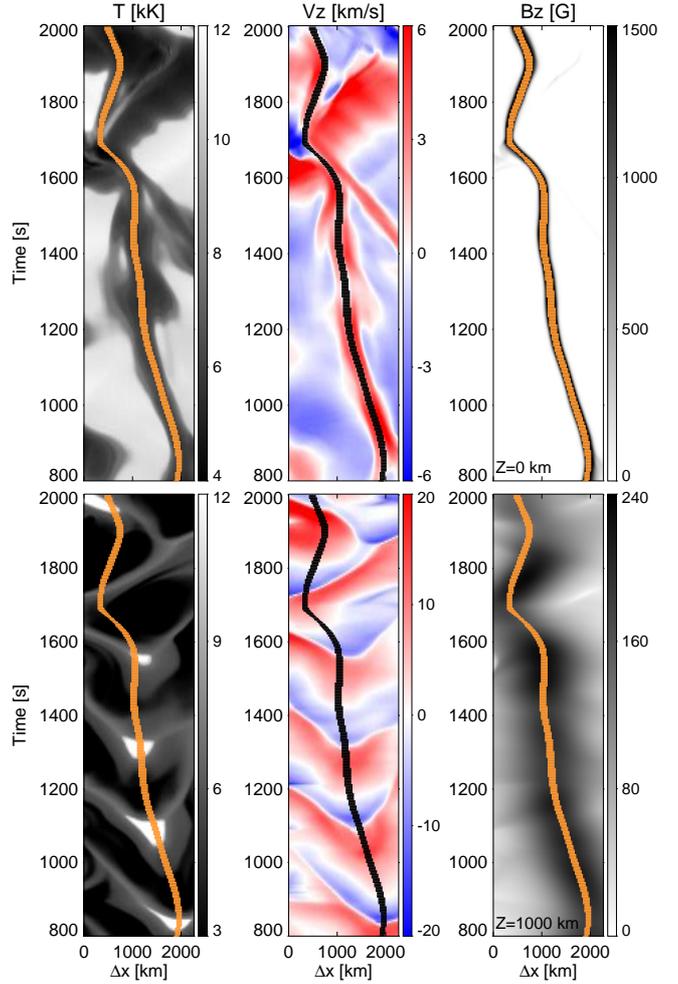}
 \vspace{+0.4cm}
 \caption{Orange (or black) colour depicts the pixels we consider belonging to the flux tube. We plot these locations over the temperature (left column), LOS velocity (middle), and longitudinal field (right) for selected heights, $z=0$~km (top row) and $z=1000$~km (bottom). The abscissa designates the horizontal computational domain while the ordinate represents time.}
 \label{Pixels}
 \end{center}
\end{figure}

\begin{figure*}
\begin{center} 
 \includegraphics[trim=0 0 15 0,width=17.0cm]{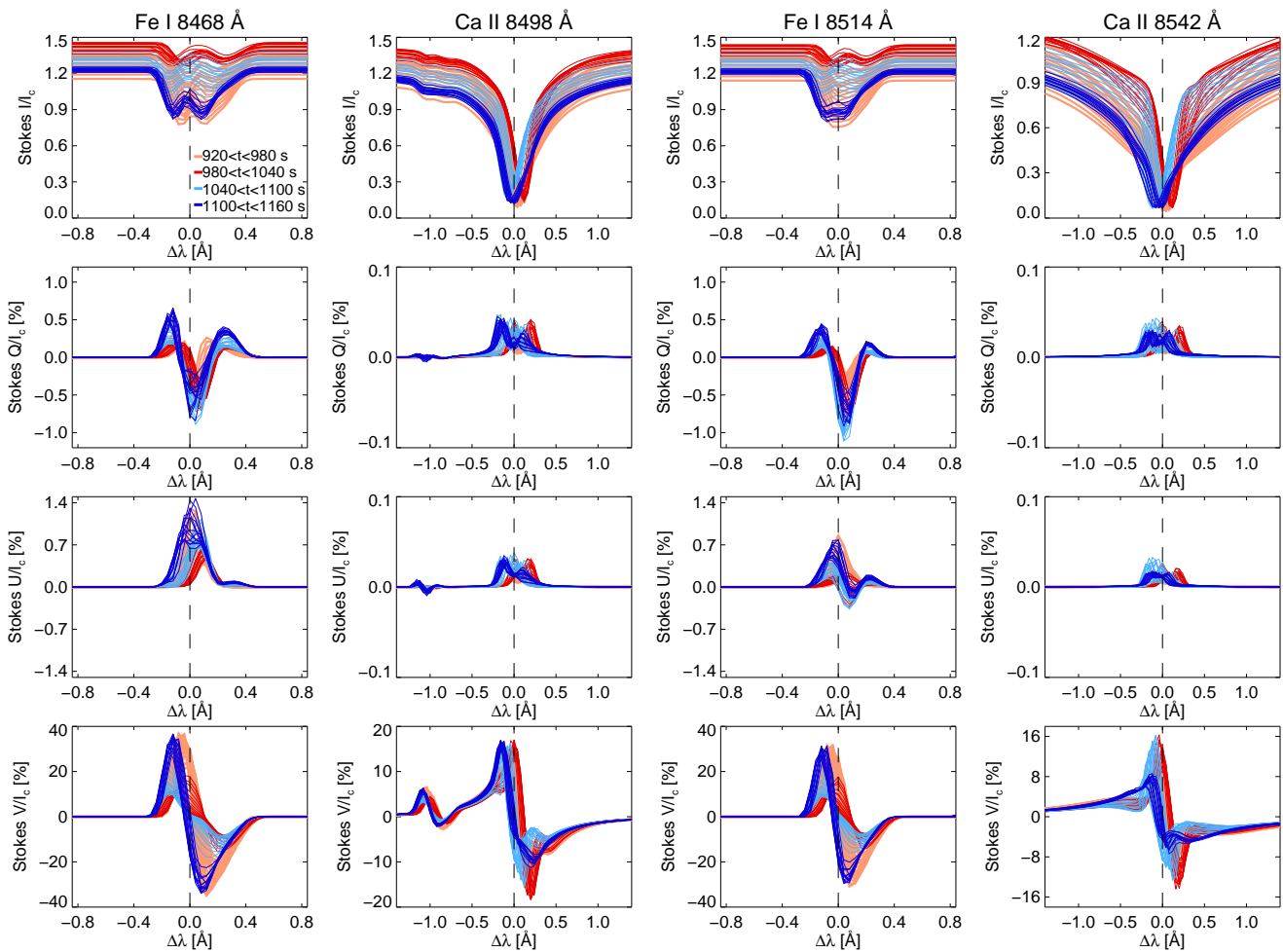}
 \vspace{+0.1cm}
 \caption{Stokes profiles inside the flux tube for the period $t=[920,1160]$~s. Each column corresponds to a different spectral line while each row displays one of the four Stokes parameters. Colours designate different reference time intervals inside the mentioned time period (see the inset of leftmost top panel). We plot individual lines every 4 seconds of simulation time.}
 \label{Shifttot}
 \end{center}
\end{figure*}

We compute the mean Stokes profiles inside the flux tube to examine the evolution of selected spectral lines. The number of pixels we use depends on the given snapshot with a mean value of $5.9\pm0.58$ pixels. In order to enhance the visualization of the different spectral features, we focus on the strongest spectral lines present in the spectrum, as we did in the first paper of these series. Those lines are the Fe~{\sc i} 8468~\AA, Ca~{\sc ii} 8498~\AA, Fe~{\sc i} 8514~\AA, and Ca~{\sc ii} 8542~\AA \ \citep[see Table 1 in][for more information]{QuinteroNoda2017}. The Stokes profiles display a periodic behaviour, following the pattern seen in the bottom row of Figure \ref{Pixels} and, for simplicity, we examine only one of these cycles, comprised of  $240$~s between $t=[920,1160]$~s. Moreover, in order to facilitate the visualization of the Stokes profiles, we divide the mentioned period into four reference intervals (depicted by different colours in Figure~\ref{Shifttot}).

The first part of the cycle (orange, i.e. $t=[920,980]$~s) shows intensity Stokes profiles (first row) slightly redshifted with respect to the rest wavelength (vertical dashed line), with a moderate-high continuum value (around 1.2 of the average continuum intensity $I_c$), and a deep line core (see the Ca~{\sc ii} lines). We can relate this high continuum value inside the flux tube to the observed quiet Sun bright points, which are associated with the presence of strong magnetic field concentrations in the solar network \cite[for instance,][]{Chapman1968}. Moreover, the spectral profile is wide and asymmetric for both the photospheric and chromospheric lines, something that also has been reported for network and plage regions observations of photospheric lines \citep[][]{MartinezPillet1997}. During the second interval (red, i.e. $t=[980,1040]$~s), strong downdrafts in the close surroundings of the flux tube pump plasma inside in the downward direction \citep[see Figure~2 of][]{Kato2016} producing strongly redshifted Stokes profiles. The continuum intensity is largely enhanced for both the photospheric and chromospheric lines, while the line core-to-continuum ratio is strongly diminished for the former lines, generating an extremely swallow absorption profile. Later, during the third part of the process (sky blue, i.e. $t=[1040,1100]$~s), slightly blueshifted profiles can be detected and the line core and continuum intensity start to decrease. This period belongs to the beginning of the rebound phase explained in the previously cited work. The final phase (blue, i.e. $t=[1100,1160]$~s), is characterized by strongly blueshifted profiles whose line core and continuum intensity have strongly decreased in comparison with the second interval (red). 

We display in the second and third rows of Figure~\ref{Shifttot} the linear polarization signals. They also show the same wavelength shift pattern found in the intensity profiles, being redshifted at the beginning of the cycle and strongly blueshifted at the end, around $240$~s later. Ca~{\sc ii} polarization signals are weak, indicating that it is almost impossible to detect them with typical integration times, e.g, a noise level of $1\times10^{-3}$ of $I_c$. However, this is something we expect from the studies presented in \cite{QuinteroNoda2016,QuinteroNoda2017} where we found that there are moderate linear polarization signals produced by chromospheric lines, but they are located in the surroundings of the magnetic flux concentrations. Additionally, if we focus on the photospheric lines, we can see that the polarization signals are different between them, with the Stokes $Q$ amplitude being larger for the Fe~{\sc i} 8514~\AA \ line and the opposite for the Stokes $U$ signals. This points out that the horizontal component of the magnetic field (just $Bx$ in this case as it is a 2D simulation) is probably changing in a height scale shorter than the difference in the height of formation between these spectral lines. Although we know that the atmospheric parameters from realistic MHD simulations largely change with height in short scales, it is noteworthy that the spectral lines are sensitive to those changes, something that we could not foresee through the response functions (RF) study in \cite{QuinteroNoda2017}.

The last row of Figure~\ref{Shifttot} displays the Stokes $V$ profiles. They show the same behaviour found for the rest of the Stokes parameters, being redshifted at the beginning of the cycle and blueshifted at the end. We can see that the Stokes $V$ polarity is the same between all the lines although their amplitude differs between photospheric and chromospheric lines. In the former case, the polarization signals are higher at the beginning and at the end of the process (orange and blue colours). However, for the latter lines, the amplitude is largest in the middle of the process, i.e. red and sky blue colours (e.g., see the red lobe of Ca~{\sc ii} 8498 and 8542~\AA \ lines). The reason could be an enhancement of the field strength in the middle part of the cycle. This enhancement is large enough to set the photospheric lines in the strong field regime (the separation between Stokes $V$ lobes increases but the amplitude barely changes) while the Ca~{\sc ii} lines are still in the weak field regime (the chromospheric field is much weaker than the photospheric one).

Finally, this process is periodically repeated for the rest of the simulation and, although the Stokes parameters slightly change between cycles, the basics of the spectral features described before are maintained.

\begin{figure*}
\begin{center} 
 \includegraphics[trim=5 0 10 0,width=17.0cm]{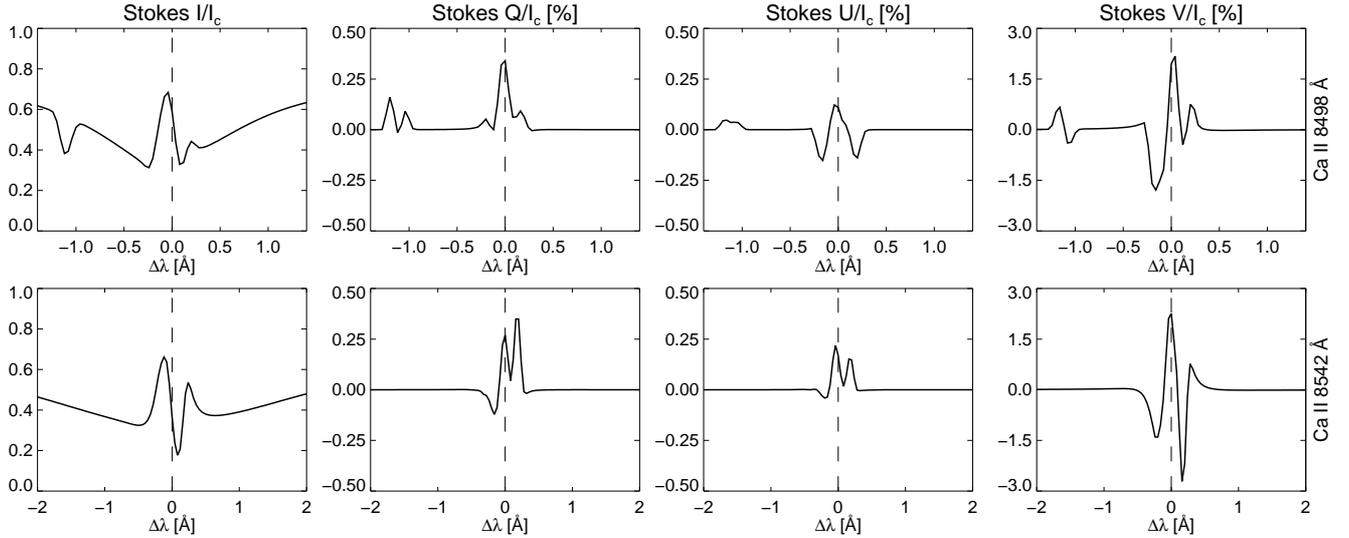}
 \vspace{+0.1cm}
 \caption{Stokes profiles at the edge of the magnetic concentration (see dashed line in Figure~\ref{Context}). Each column depicts a different Stokes parameter while rows display the profiles for the Ca~{\sc ii}~8498~\AA \ (top) and Ca~{\sc ii}~8542~\AA \ (bottom) lines.}
 \label{edgeF}
 \end{center}
\end{figure*}

\subsection{Stokes profiles at the edge of the flux tube}\label{edge}

The study presented in \cite{delaCruzRodriguez2013} shows that there are Ca~{\sc ii} 8542~\AA \ intensity profiles located at the edges of bright points that show an enhanced line core intensity. They explained that the reason for this is that those locations correspond to regions where there is a strong magnetic field gradient along the line of sight, i.e. a weak magnetic field in the photosphere and stronger in the chromosphere due to the presence of a magnetic canopy. Moreover, this magnetic canopy also produces a very steep temperature rise. The authors emphasized the importance of these particular profiles as they could indeed be related to some mechanism that transfers energy to upper heights, e.g. there is a certain probability of the appearance of currents (and associated dissipation) in the vicinity of flux tube concentrations.

In this work, we have the opportunity to deepen into this topic as we can study not only the polarization signals produced by the longitudinal component of the magnetic field, but also the linear polarization signals generated by the transversal component. In particular, we aim to examine the amplitude of these signals. For this purpose, we show in Figure~\ref{edgeF} the Stokes profiles at the edge of the magnetic concentration (see dashed line in Figure~\ref{Context}) for the chromospheric lines. The intensity profiles display the characteristic behaviour mentioned in \cite{delaCruzRodriguez2013} with a raised core and two emission lobes at both sides of the line centre. The intensity of these emission lobes, created by LOS velocity gradients, is asymmetric for both calcium lines. 

In the case of the linear polarization profiles, we can see moderate signals of almost $4\times10^{-3}$ of $I_c$ while the amplitude of the circular polarization is much larger reaching values of up to $3\times10^{-2}$ of $I_c$. This means that the magnetic field at the height of sensitivity of the Ca~{\sc ii} lines has a certain degree of inclination with respect to the line of sight. This is because the magnetic field expands with height and occupy larger areas, in this case at more than 400~km from the centre of the magnetic flux concentration. Finally, these results indicate that we would be able to detect the polarization signals produced at the edge of the flux tube with high spatial resolution observations and with a low polarimetric noise level, e.g. $5\times10^{-4}$ of $I_c$.

\begin{figure*}
\begin{center} 
 \includegraphics[trim=-10 0 7 0,width=17.2cm]{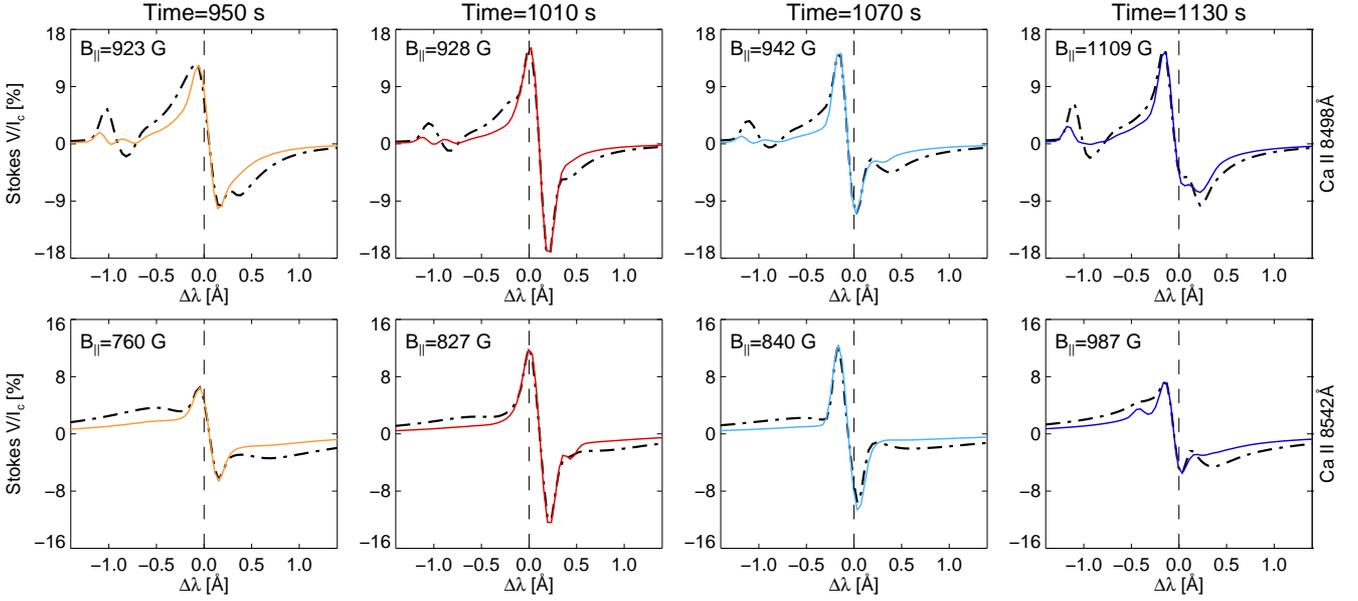}
 \vspace{+0.05cm}
 \caption{Circular polarization profiles inside the flux tube. Stokes $V$ spectra are displayed with black dotted lines while $-\Delta\lambda_B\frac{\partial I}{\partial \lambda}$ is depicted in solid colour. Each column corresponds to a given instance (see also Figure~\ref{Shifttot}) while rows show the results for the Ca~{\sc ii} 8498~\AA \ (top) and Ca~{\sc ii} 8542~\AA \ (bottom) lines. We include the magnetic field $B_{\parallel}$ used to fit the left lobe of Stokes $V$ inside each panel.}
 \label{deltai}
 \end{center}
\end{figure*}

\subsection{Magnetic field determination based on the weak field approximation}\label{RFstudy}

The results presented in the previous section show that the amplitude of the Stokes profiles varies with time (see Figure~\ref{Shifttot}). There are multiple factors that can induce this: one could be just a change of the magnetic field strength, a second option could be a temperature enhancement, and a third one, a shift in height of formation due to a Wilson depression effect \cite[for instance,][]{Bray1964}. If we examine the scatter plots presented in Figure~3 of \cite{Kato2016}  we can see that the longitudinal field changes only slightly with time, while the temperature shows a large variation, mainly above 400~km. This means that the large fluctuations found in the Stokes $V$ parameters are not mainly due to changes in the field strength.

In order to shed light on this, we follow a very simple approach commonly used in spectropolarimetric observations. We know that Stokes $V$ is directly related to Stokes $I$ in the weak field regime \citep[see chapter 9 in the monograph of][for more details]{Landi2004}, being proportional to its derivative following the relation
\begin{equation}
V(\lambda)=-\Delta\lambda_B\frac{\partial I}{\partial \lambda}
\end{equation}\label{weakfield}
where
\begin{equation}
\Delta\lambda_B=4.67\times10^{-13}g_{\rm eff}\lambda^2B_{\parallel},
\end{equation}
with $g_{\rm eff}$ the effective Land\'{e} factor, $\lambda$ the wavelength used as a reference in \AA, $B_{\parallel}$ the longitudinal component of the magnetic field (parallel to the observer's line of sight) in~G, and $\Delta\lambda_B$ the factor we need to apply to the derivative $\frac{\partial I}{\partial \lambda}$ in order to match the Stokes $V$ amplitude. We aim to use this approximation with the two Ca~{\sc ii} infrared lines as, in general, they form under the weak field regime \citep[additional information can be found in, for instance, the review of][]{delaCruzRodriguez2016}. We exclude from this study the two photospheric lines because they show indications of being under the strong field regime, i.e. the separation between lobes ($\sigma-$components) changes with time. This can be seen in the red and sky-blue profiles in Figure~\ref{Shifttot}. 

\begin{figure}
\begin{center} 
 \includegraphics[trim=-5 0 10 0,width=8.0cm]{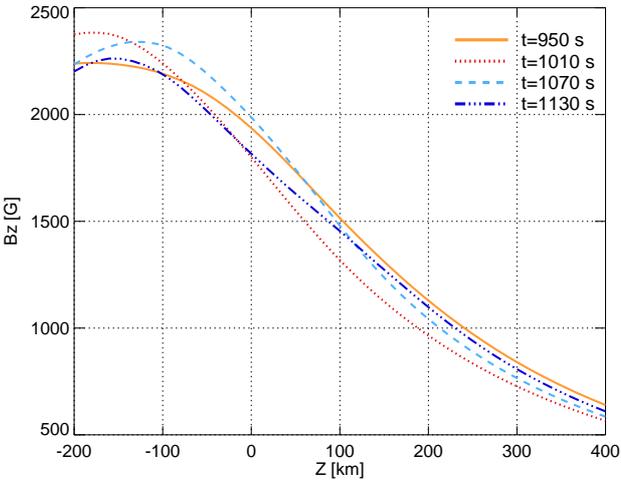}
 \vspace{-0.2cm}
 \caption{Height stratification of the longitudinal component of the magnetic field extracted from the 2D numerical simulation (e.g., see Figure~\ref{Context}) for four selected instances (see Figure~\ref{deltai}).}
 \label{fieldstrength}
 \end{center}
\end{figure}

We show in Figure \ref{deltai} the results for four selected instants. They correspond to the central time of the four reference intervals used in Figure~\ref{Shifttot}. We adjust $B_{\parallel}$ in order to match the Stokes $V$ left lobe (solid lines). We show in each panel the results of the inferred magnetic field strength ($B_{\parallel}$) that has a mean value, and a standard deviation, for the four selected snapshots of $976\pm89$~G and $853\pm95$~G for Ca~{\sc ii} 8498 and 8542~\AA \ lines, respectively.  If we compare these results with the longitudinal field component of the simulation (see Figure~\ref{fieldstrength}), we can see that they correspond to the field strength at heights around $200\sim300$~km, what roughly indicates that the line core of the Ca~{\sc ii} lines is sensitive to the magnetic field at those heights. In order words, around 500~km above $\tau=1$, if we take into account that, inside the flux tube, the mean height where the continuum optical depth is unity is -200~km \citep{Kato2016}. 

\begin{figure*}
\begin{center} 
 \includegraphics[trim=7 0 0 0,width=16.1cm]{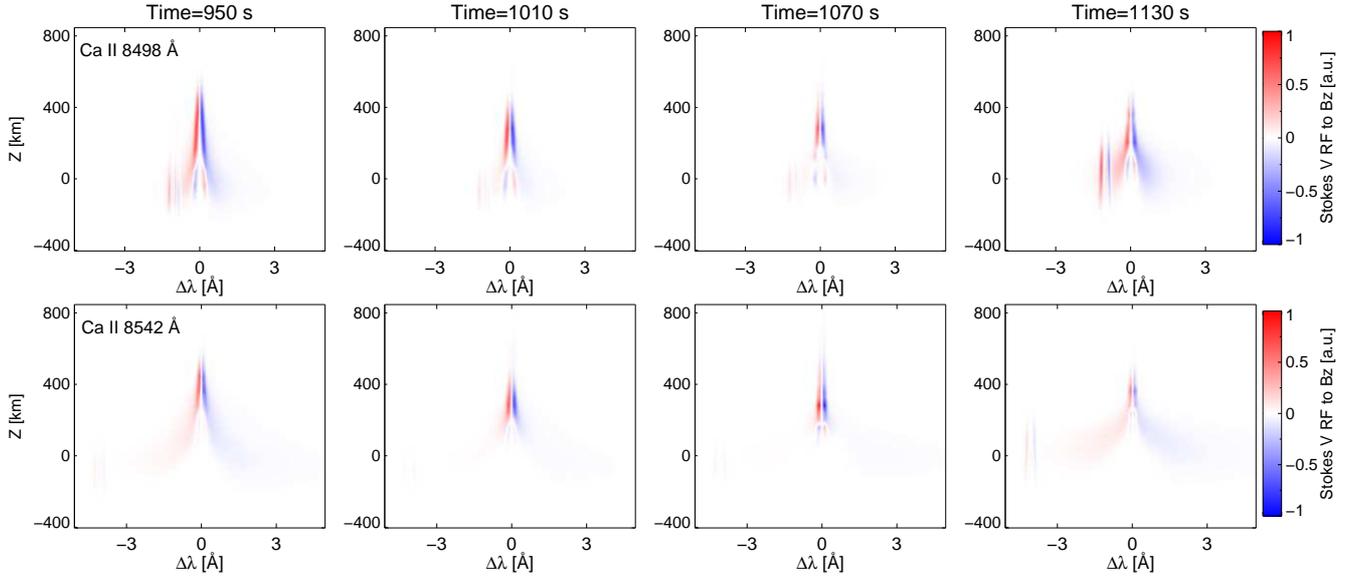}
 \vspace{+0.1cm}
 \caption{Stokes $V$ response functions to perturbations in the longitudinal field for selected time instances (columns), see also Figure~\ref{deltai}. First row shows the results for the Ca~{\sc ii} 8498 \AA \ line while bottom row corresponds to the Ca~{\sc ii} 8542 \AA \ line. RF are normalized to their individual maximum value.}
 \label{2dRF}
 \end{center}
\end{figure*}

Regarding the evolution of the inferred field strength values, $B_{\parallel}$ slightly increases for the first three instances and there is a larger jump in the last one. If we compare these results with the longitudinal component of the magnetic field from the 2D simulation (see Figure~\ref{fieldstrength}), we can see that the field strength changes with time, but these variations do not reflect the weak field approximation results. At upper heights, the largest field values correspond to the first snapshot, while the lowest ones are associated with the second instant. Therefore, it seems that the evolution of the Stokes profiles is also proving the changes in the height of formation of the Ca~{\sc ii} lines. This means that, as the inferred magnetic field for the last instance (dashed-dotted blue) is larger than the one from the first time (solid orange), something that never happens for heights above -100~km, the height where the line forms at that time should be lower than that from the first instant. It is also possible that the opposite happens, or even a combination of height shifts in time. 

We plan to check this in the following section computing the response functions (RF) to changes in the magnetic field. However, before that, we would like to mention that, in all the examined cases, -$\frac{\partial I}{\partial \lambda}$ accurately matches the Stokes $V$ profiles. This indicates that the changes in Stokes $V$ in this scenario are mainly due to variations in the intensity profiles. Something that is in some sense in agreement with the work of \cite{delaCruzRodriguez2013} where they reached a similar conclusion using spectropolarimetric observations, i.e. the change of polarity on Stokes $V$ that appeared during the umbral flash is well fitted with the weak field approximation and an almost constant magnetic field.

Finally, concerning the differences between the two Ca~{\sc ii} lines, the inferred magnetic field values are always larger for the Ca~{\sc ii} 8498~\AA \ line. As the magnetic field strength monotonically decreases with height (see Figure~\ref{fieldstrength}), this indicates that the Ca~{\sc ii} 8498~\AA \ line indeed forms in lower atmospheric layers as its smaller $gf$ value (by a factor of 9 compared to the 8542~\AA \ line) would already suggest.

\subsection{Response functions to perturbations in $B_z$}

Previous results indicate that the height of formation of the Ca~{\sc ii} lines varies during the magnetic pumping process, i.e. the field strength inferred from the weak field approximation does not exactly correspond with the evolution of the longitudinal component of the simulated magnetic field at a given height. Therefore, we check in this section whether these results are due to some limitation of the weak field approximation or if, in fact, the height where the line is sensitive to the field strength fluctuates with time.

\begin{figure}
\begin{center} 
 \includegraphics[trim=0 0 12 0,width=7.5cm]{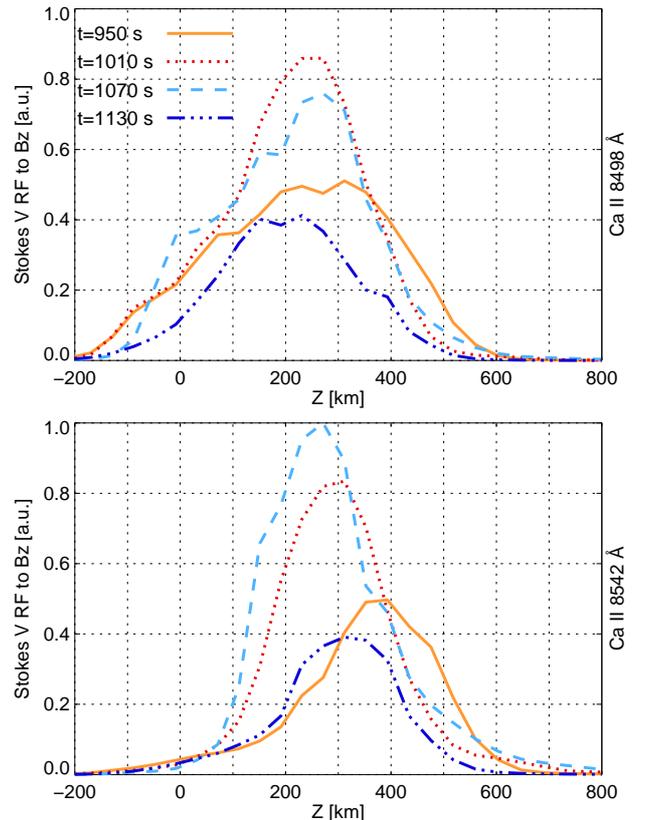}
 \vspace{-0.0cm}
 \caption{Maximum of the absolute value of the Stokes $V$ RF to perturbations on $B_z$ for the line core wavelengths,  i.e. $\lambda_c - 0.6$~\AA \ $\leq$  $\lambda$ $\leq \lambda_c +0.6$~\AA. Top panel displays the results for the Ca~{\sc ii} 8498~\AA \ line while the bottom panel corresponds to the Ca~{\sc ii} 8542~\AA \ line. RF are normalized to the maximum value of the Ca~{\sc ii} 8542~\AA \ RF at t=1070~s. Line-style colour code is the same as was used in Figure~\ref{fieldstrength}.} 
 \label{Comparheightb}
 \end{center}
\end{figure}

\begin{figure*}
\begin{center} 
 \includegraphics[trim=0 0 5 0,width=15.0cm]{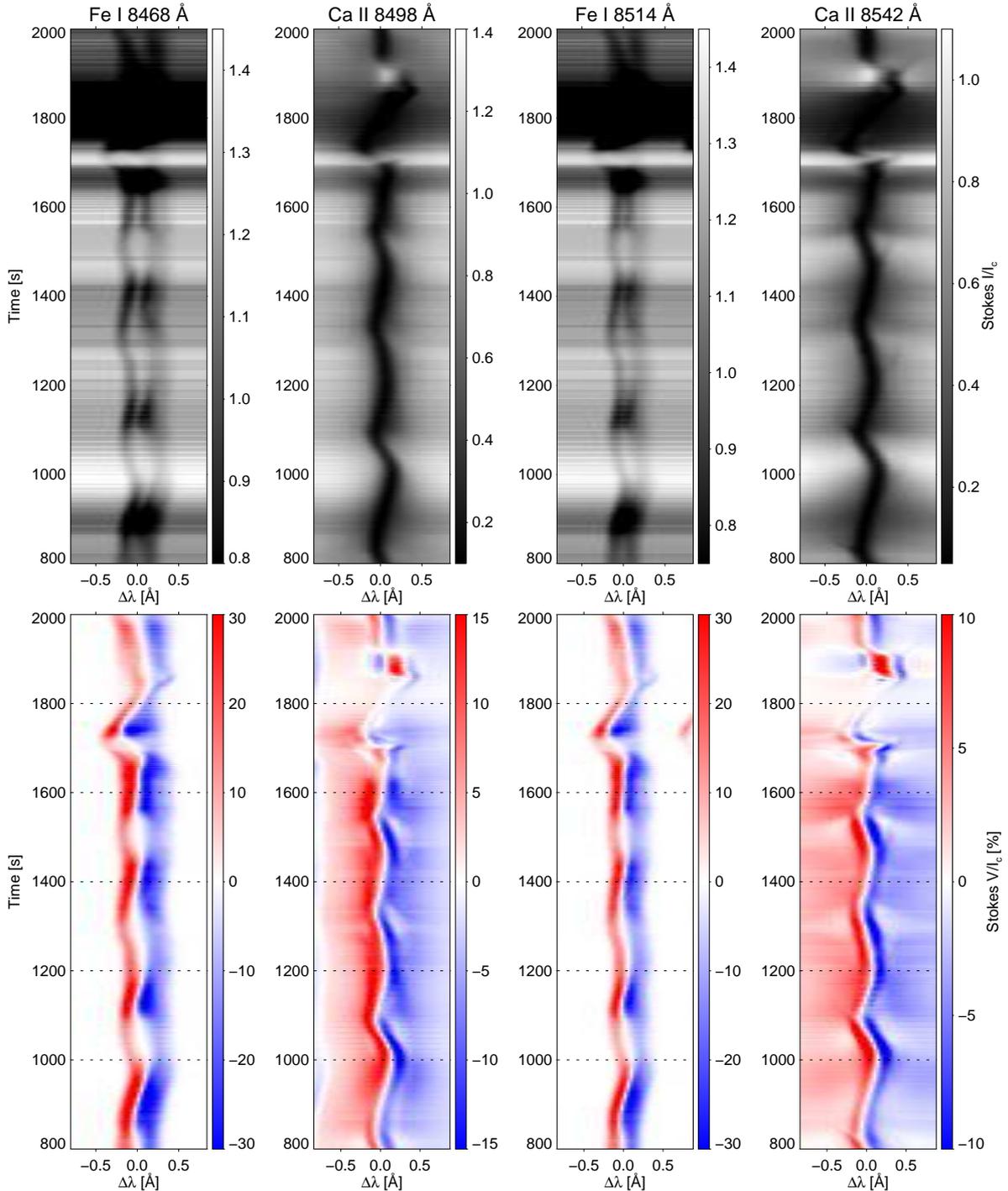}
 \vspace{-0.0cm}
 \caption{Computed mean Stokes $I$ (top) and $V$ (bottom) profiles using the pixels highlighted in Figure~\ref{Pixels}. We plot them as a function of wavelength and time for the whole simulation run used in this work.}
 \label{Lambdat}
 \end{center}
\end{figure*}

\begin{figure*}
\begin{center} 
 \includegraphics[trim=10 0 10 0,width=15.5cm]{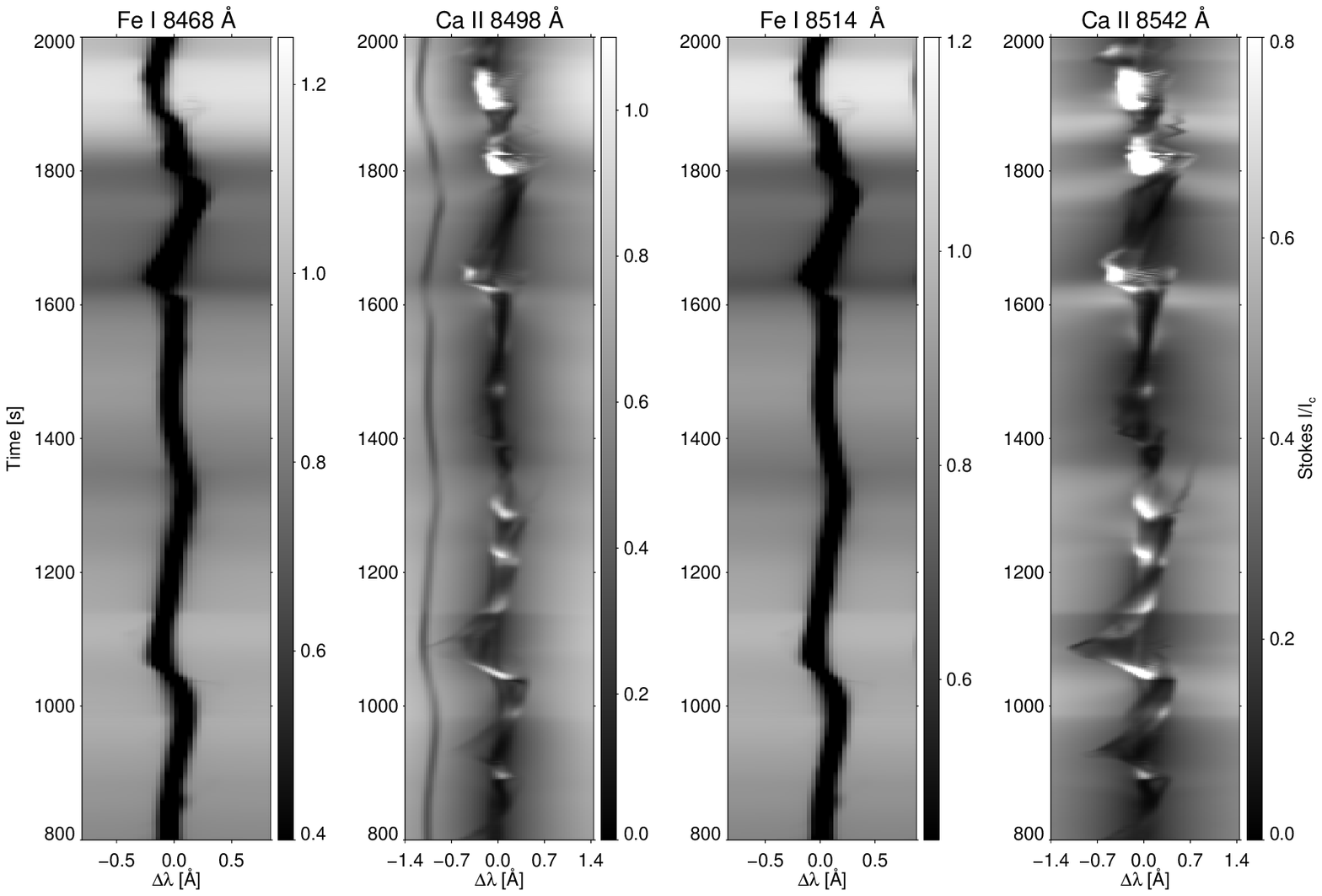}
 \vspace{-0.0cm}
 \caption{Stokes~$I$ profiles using pixels that are located approximately at 3.5~Mm from the flux tube concentration. We compute the mean profiles using 6 pixels to simulate the average number of pixels used in previous sections. They are plotted as a function of wavelength and time for the whole simulation run used in this work.}
 \label{Lambdatout}
 \end{center}
\end{figure*}

We compute the Stokes $V$ response function (RF) \citep[for instance,][]{Landi1977} to changes in the longitudinal component of the magnetic field following the method explained in \cite{QuinteroNoda2016}, using a perturbation value of $\Delta B_z=1$~G. Figure~\ref{2dRF} shows the results for the same time instances studied before. There is a strong contribution from the line core wavelengths of both lines that is located in the height range  $200-500$~km for the whole period. It seems that the height of maximum sensitivity is larger at the beginning and lower in the last step. In order to confirm this, we plot in Figure~\ref{Comparheightb} the maximum of the absolute value of the Stokes $V$ RF for the wavelength range $\lambda_c - 0.6$~\AA \ $\leq$  $\lambda$ $\leq \lambda_c +0.6$~\AA \ (with $\lambda_c$ the line core wavelength), at a given height for both Ca~{\sc ii} lines. This plot is similar to that of Figure~4 in \cite{QuinteroNoda2017} and allows the visualization of the RF in a one dimensional format. We can see that, in general, the results are similar to those found in the mentioned work where the Ca~{\sc ii} 8498~\AA \ forms slightly lower in the atmosphere. Moreover, in all cases, the Ca~{\sc ii} 8498~\AA \ RF extends to deeper layers indicating that is more sensitive to the low photosphere. On the other hand, the RF of both lines quickly drops after 600~km (800~km above $\tau=1$ inside the flux tube). 

Finally, if we examine the evolution of the height where the RF are maximum, we can see that it fluctuates with time between approximately 200 and 400~km. Moreover, it is highest for the first instant (solid orange), which explains why the inferred field strength values (see Figure~\ref{deltai}) are smaller for the first instant (see also solid orange in Figure~\ref{fieldstrength}). Thus, although the magnetic field is stronger at this time for any given height above 100~km, the formation height of the lines is higher and they are sensitive to an ``effective'' magnetic field that is weaker. The opposite is happening in the last instance, where the inferred magnetic field is stronger because the lines are sensitive to lower heights (dashed-triple dot blue), and, therefore, to an ``effective'' magnetic field that is stronger.

\subsection{Evolution of polarization signals}

Figure~\ref{Lambdat} displays the evolution of the mean Stokes~$I$ and $V$ profiles inside the flux tube for the four spectral regions studied before. The intensity profiles (top row) shows the aforementioned oscillatory pattern in all the spectral lines and the presence of periodic large Zeeman splitting in the Fe~{\sc i} 8468~\AA \ and 8514~\AA \ lines. We can see in the bottom row a similar behaviour for the circular polarization signals (we leave out of this study the linear polarization signals because, in general, they are weak inside the flux tube). Moreover, there is an area asymmetry in the Stokes~$V$ profiles with the red lobe always being broader \citep[see also][]{MartinezPillet1997}. 

If we compare these results with previous studies of chromospheric oscillations, we find that they resemble those presented in \cite{Carlsson1997}, based on a one-dimensional simulation using a piston perturbation. In addition, there are similar examples with polarimetric observations and non-LTE synthesis in \cite{Pietarila2007} and \cite{delaCruzRodriguez2013}. In the former work, the line core of Ca~{\sc ii} lines often shows large shifts that correspond to acoustic shocks. We find similar shifts for the Ca~{\sc ii} lines (and to a lesser extent for the photospheric lines), e.g. $t=$ [830, 1100, 1550]~s.

In addition, there is a large oscillation and swaying motion at the end of the period (around $t=$ 1700~s) that strongly affects the Stokes profiles. This event was defined in \cite{Kato2016} as a rapid and large change of transverse velocity amplitude. The authors explained (see Appendix B of their study) that this event is produced by a shock that propagates in the horizontal domain of the simulation box. During this period, photospheric lines display a traditional absorption profile but chromospheric lines are very complex. Their line core is in absorption but their \textit{knees} are in emission with an intensity closer to the continuum value and different amplitude between the blue and the red component.

\subsection{Quiet Sun regions}

We evaluate in this section whether the oscillations of the atmospheric parameters found inside the flux tube are also present outside the magnetic concentration and whether they produce any characteristic feature in the Stokes profiles. Thus, we compute the mean Stokes profiles at around 3.5~Mm from the flux tube (see Figure~\ref{Context}) and we plot the results in Figure~\ref{Lambdatout}. We only show the intensity profiles because, outside the flux tube concentration, the polarization signals are low, i.e. around $3\times10^{-4}$ of $I_c$ for the Ca~{\sc ii} lines. We can see that the chromospheric lines sometimes show large dopplershifts of more than 30~km/s (for instance $t\sim1100$~s). Moreover, the absorption profile of the Ca~{\sc ii} lines sometimes displays a complex pattern with one of its \textit{knees} in emission and the other in absorption. This is something that has been already reported by \cite{delaCruzRodriguez2013} \citep[see also Figure~10 in][]{delaCruzRodriguez2016} where the authors explained that a rise in the temperature of upper layers (probably due to a shock) induces the enhanced line core emission. Some of those large dopplershifts can be detected in the photospheric lines although with much less amplitude. Moreover, we can see that they are always in absorption and there is no trace of Zeeman splitting.

\subsection{Extreme dopplershift}

Blended lines that form at different geometrical heights can hinder or distort the atmospheric information as the spectral features from one spectral line can get mixed with a second line or more blended lines. In \cite{QuinteroNoda2017} we mentioned that, contrary to the case of the Ca~{\sc ii} 8662~\AA \ line, the  Ca~{\sc ii} 8498~\AA \ line core is free of blends. However, there is a photospheric line, i.e. the Fe~{\sc i} 8497~\AA, located at 1~\AA \ from its line core. In that work, we explained that this does not pose any disadvantage, except when there is an extreme upflow at chromospheric layers that is not present in the photosphere (where the Fe~{\sc i} 8497~\AA \ forms). 

In the previous section, we examined a region outside the flux tube where strong dopplershifted signals were detected for chromospheric lines. In particular, we found that around $t=1100$~s there is an extreme blueshift that displaces the line core of the Ca~{\sc ii} 8498~\AA \ to very close to the rest wavelength of the Fe~{\sc i} 8497~\AA \  photospheric line. Therefore, in this section, we study the intensity profiles for this period of time to check if the Ca~{\sc ii} 8498~\AA \ line core gets fused together with the photospheric line, hindering the analysis of those pixels.

\begin{figure}
\begin{center} 
 \includegraphics[trim=10 0 10 0,width=8.5cm]{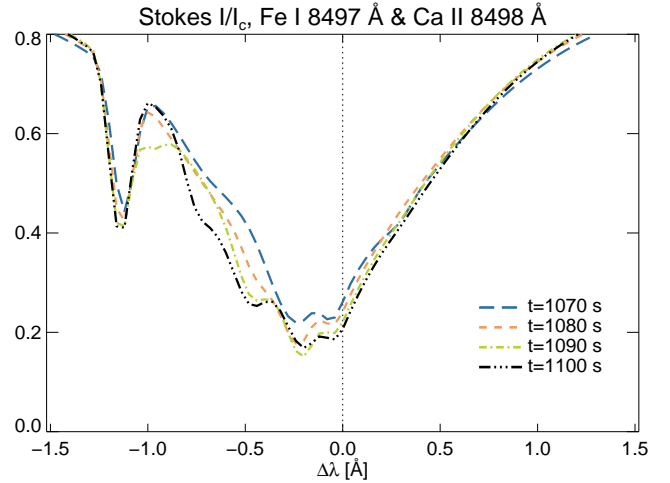}
 \vspace{-0.5cm}
 \caption{Fe~{\sc i} 8497~\AA \ and Ca~{\sc ii} 8498~\AA \ intensity profiles from a region that presents high velocities at chromospheric layers (see Figure~\ref{Lambdatout}). Each linestyle corresponds to a different time step starting at $t=1070$~s (dashed blue) and finishing at $t=1100$~s (dashed-triple dot black). These profiles correspond to a region located at 3.5~Mm from the flux tube concentration.}
 \label{Extreme}
 \end{center}
\end{figure}

Figure \ref{Extreme} shows Stokes $I$ for the mentioned interval of time for both spectral lines. At the beginning of the selected instance, i.e. $t=1070$~s, the Ca~{\sc ii} line is slightly blueshifted and it can be perfectly distinguished from the photospheric line. As the simulation evolves, the chromospheric line becomes wider and strongly blueshifted ($t=1090$~s). However, even in that case, the Ca~{\sc ii} line core does not reach the photospheric line, being the latter always clearly visible during the whole period. Therefore, our previous statement that the blended Fe~{\sc i} 8497~\AA \ line is located far enough to not disturb the line core of the Ca~{\sc ii} 8498~\AA \ line, is still correct.

\section{Concluding remarks}

We examined in this work the two-dimensional simulation presented in \cite{Kato2016}. The atmosphere inside the flux tube periodically changes due to the magnetic pumping process and the acoustic waves generated after it. This variation of the atmospheric properties leaves a characteristic imprint in the Stokes profiles of photospheric and chromospheric spectral lines through large dopplershifts and Stokes~$I$ line core and Stokes $V$ amplitude fluctuations. We also studied the polarization signals at the edge of the magnetic concentration concluding that they can be detected with a noise level lower than $1\times10^{-3}$ of $I_c$. Later, we examined in more detail the Stokes~$V$ amplitude changes with time analysing the results of the weak field approximation and the Stokes $V$ RF to perturbations on the longitudinal component of the magnetic field. We concluded that the magnetic pumping process modifies the height of formation of the Ca~{\sc ii} lines which causes the inferred magnetic field strength to deviate from the expected values. These results indicate that the changes in the Stokes $V$ profiles come from several sources: the inherent field strength evolution, temperature variations, and also fluctuations of the height of formation. Therefore, for the current simulation conditions, determining the evolution of the magnetic field strength, disentangling the above effects, is difficult. Moreover, this is something that is not related to the weak field approximation, as it is an effect also found in the Stokes $V$ RF to $Bz$. Therefore, we also expect the same uncertainties when we perform inversions of the Stokes profiles.

We additionally studied a different spatial region outside the flux tube concentration to check whether the characteristics of the atmosphere were different, and whether this property leaves an imprint in the intensity profiles. We found a much more dynamic and complex atmosphere where extreme dopplershifts occur mainly in the chromosphere. The amplitude of those wavelength shifts impelled us to check if the Ca~{\sc ii} 8498~\AA \ line core could be mixed with the iron line that is blended with its blue wing. Fortunately, we found that the wavelength shift is not high enough to prevent the useful analysis of both lines independently.

The results of the present work demonstrate that by observing the spectral lines belonging to the 850~nm window, we are sensitive to the physical phenomenon, i.e. magnetic pumping, that takes place in the simulation from the photosphere to the chromosphere. However, we recommend special care when inferring the field strength, through the weak field approximation or inversions of the Stokes profiles, on solar observations that resemble the present simulation, as the results would be inaccurate. We cannot provide an error value at this moment, although we aim to do it in the future through non-local thermodynamic equilibrium inversions of the synthetic spectra, including photospheric and chromospheric lines. 

Finally, gathering all the results presented here, including the previously mentioned limitations, we reiterate that observing the 850~nm window allows covering a large range of atmospheric layers with just a single spectral channel. Therefore, we believe that there is no better option in the visible-infrared spectrum that can provide the same information.

\section*{Acknowledgements}
Special thanks to Sami Solanki and Valent\'{i}n Mart\'{i}nez Pillet for their comments and suggestions. This work was supported by the funding for the international collaboration mission (SUNRISE-3) of ISAS/JAXA. The simulation data leading to these results have received support by the Research Council of Norway, grants 221767/F20 and 2309038/F50, and through grants of computing time from the programme for supercomputing. JdlCR is supported by grants from the Swedish Research Council (VR) and the Swedish National Space Board (SNSB). The research leading to these results has received funding from the European Research Council under the European Union's Seventh Framework Programme (FP7/2007-2013)/ERC grant agreement no 291058. This work has also been supported by Spanish Ministry of Economy and Competitiveness through the project ESP-2016-77548-C5-1-R.

\bibliographystyle{mnras} 
\bibliography{multiline2} 

\bsp	
\label{lastpage}
\end{document}